\documentclass[aps,prl,twocolumn,superscriptaddress,nofootinbib,floatfix]{revtex4-2}

\usepackage[T1]{fontenc}
\usepackage{amsmath,amssymb,bm}
\usepackage{mathrsfs}
\usepackage{hyperref}

\newcommand{\Ad}{\operatorname{Ad}}
\newcommand{\ad}{\operatorname{ad}}
\newcommand{\Tr}{\operatorname{Tr}}
\newcommand{\tr}{\operatorname{tr}}
\newcommand{\dd}{\mathrm{d}}
\newcommand{\ess}{\operatorname*{ess\,inf}}
\newcommand{\avg}[1]{\langle #1\rangle}
\newcommand{\ip}[2]{\langle #1,#2\rangle}

\begin{document}

\title{Curvature-Orbit Quantization of Gauge-Related Connections\\
and a Radon--Nikodym Route to Lattice Yang--Mills Theory}

\author{Victor Atanasov}
\email{vatanaso@phys.uni-sofia.bg}
\affiliation{Condensed Matter Physics and Microelectronics Department,
Faculty of Physics, Sofia University ``St.\ Kliment Ohridski'',
5 J.\ Bourchier Blvd., 1164 Sofia, Bulgaria}

\begin{abstract}
We study a four-dimensional gauge construction in which two gauge-related
non-abelian connections, $A=B^g$, interact only through the squared difference
of their curvatures. Independent kinetic term is absent. The action
is an ultralocal functional on a curvature orbit. Its Hessian
is a pointwise positive-semidefinite operator whose kernel is the common
stabilizer algebra of the curvature components, and its spectrum is given for every compact simple Lie group. At a finite regulator the model is an exact compact-group integral. Integrating out the orbit variable yields a positive, bounded
Radon--Nikodym cocycle that relates the induced link measure to a lattice
Yang--Mills measure, recasting the existence and mass-gap question as a uniform estimate on this cocycle. 
\end{abstract}

\maketitle

\emph{Introduction.}---The Yang--Mills existence and mass-gap problem asks for
a mathematically complete four-dimensional quantum gauge theory with a strictly
positive spectral gap~\cite{JW}. Lattice regularization~\cite{Wilson,Creutz}
supplies finite-volume measures with a rigorous Osterwalder--Schrader
structure~\cite{OS1,OS2,OSeiler}, but the continuum limit and the gap remain
open. Here we examine a different starting point: two connections related by a
gauge transformation, coupled only through the difference of their curvatures.
The construction was motivated by the curvature-difference formalism used for
gravitational waves beyond the linear approximation~\cite{victor}.

Our results are the following. (i) The action is an exact, ultralocal orbit
functional whose zero set is the curvature stabilizer. (ii) Its Hessian is a
pointwise positive-semidefinite operator with kernel equal to the stabilizer
algebra, and its spectrum is explicit and algebraic. (iii) The finite-regulator
partition function is an exact compact-group integral with exactly known first
and second cumulants and rigorous two-sided bounds. (iv) Integrating out the
orbit variable produces a positive, bounded Radon--Nikodym (RN) cocycle
relating the resulting lattice measure to a Yang--Mills measure. (v) A winding
number does \emph{not} follow from the action. We make no claim to have constructed a continuum limit or proved a
mass gap; a dedicated paragraph below separates what is known from what is open.

\emph{Curvature-displacement action and curvature orbit.}---Let
$A=B^g=g^{-1}Bg+g^{-1}\dd g$ be gauge-related connections on a principal
$SU(3)$ bundle over a Euclidean four-manifold $M$ (anti-Hermitian
conventions), so that $F_A=g^{-1}F_Bg$. With $\Delta F=F_A-F_B$ the action is
\begin{align}
\nonumber S[g;B]&=-\!\int_M\!\dd^4x\,\Tr(\Delta F_{\mu\nu}\Delta F^{\mu\nu})\\
  &=2\!\int_M\!\dd^4x\,V_B(x,g),\label{eq:S}\\
 V_B&=\sum_{\mu<\nu}\bigl\|\Ad_{g^{-1}}F_{B,\mu\nu}-F_{B,\mu\nu}\bigr\|^2,
\label{eq:V}
\end{align}
with the positive invariant inner product $\ip XY=-\Tr(XY)$ on
$\mathfrak{su}(3)$. The functional depends on $g(x)$ only through
$\Ad_{g(x)^{-1}}F_B(x)$: it is \emph{ultralocal} in $g$.

At each point define the common curvature stabilizer and its algebra,
\begin{align}
 H_B(x)&=\{h\in SU(3):\Ad_{h^{-1}}F_{B,\mu\nu}(x)=F_{B,\mu\nu}(x)\nonumber\\
 &\qquad\qquad\qquad\qquad\forall\,\mu<\nu\},\nonumber\\
 \mathfrak h_B(x)&=\textstyle\bigcap_{\mu<\nu}\ker\ad_{F_{B,\mu\nu}(x)} .
\label{eq:stab}
\end{align}
Then $S=0$ if and only if $g(x)\in H_B(x)$ almost everywhere, and the
physical variable is the point of the \emph{curvature orbit}
$H_B(x)\backslash SU(3)$, of dimension $8-\dim\mathfrak h_B(x)$: six for a
regular Cartan configuration (stabilizer $T^2$), eight for a generic
noncommuting one (stabilizer $\subseteq Z_3$). The connection stabilizer
$\mathcal G_B=\{g:B^g=B\}$ is contained in the curvature stabilizer, and the
inclusion may be strict. The stabilizer dimension can jump on $M$, so the
global orbit structure is stratified in general.

The stationarity conditions follow from $g_t=ge^{t\xi}$ and $B_t=B+ta$:
\begin{equation}
 \sum_{\mu<\nu}[F_{B^g,\mu\nu},F_{B,\mu\nu}]=0,\qquad
 D_B^\mu\bigl[(1-\mathcal K_g)F_{B,\mu\nu}\bigr]=0,
\label{eq:eom}
\end{equation}
with $\mathcal K_g=\tfrac12(\Ad_g+\Ad_{g^{-1}})$. The first states that the
curvature displacement is orthogonal to the tangent space of the orbit.

\emph{Hessian and stabilizer.}---Near a zero-action point write $g=e^{\epsilon}$
with $\epsilon\in\mathfrak{su}(3)$. Since $\Ad_{g^{-1}}=e^{-\ad_\epsilon}$,
\begin{equation}
 S^{(2)}[\epsilon;B]=2\!\int_M\!\dd^4x\,\ip{\epsilon}{K_B\epsilon},\;
 K_B=-\!\!\sum_{\mu<\nu}\!\ad^2_{F_{B,\mu\nu}},
\label{eq:K}
\end{equation}
with neglected terms $O(\epsilon^4)$ (the cubic term vanishes identically by
skew-adjointness of $\ad_Y$. Because
$\ad_F$ is skew-adjoint for the invariant form,
\begin{align}
 \ip{\epsilon}{K_B\epsilon}&=\sum_{\mu<\nu}\|[F_{B,\mu\nu},\epsilon]\|^2\ge0,\nonumber\\
 \ker K_B(x)&=\mathfrak h_B(x).
\label{eq:pos}
\end{align}
$K_B$ is a pointwise $8\times8$ multiplication operator: it contains no
derivative, so the quadratic theory has no covariant Laplacian and no
propagation. For a general compact group the same holds verbatim, and
\begin{equation}
 \tr_{\mathfrak g}K_B=2h^\vee\sum_{\mu<\nu}\|F_{B,\mu\nu}\|^2,
\label{eq:sumrule}
\end{equation}
where $h^\vee$ is the dual Coxeter number. (At a stationary point with nonzero
displacement the Hessian contains an additional curvature-mismatch term of
indefinite sign, so such points need not be minima.)

\emph{Explicit spectrum.}---Let all curvature components be Cartan valued,
$F_{B,\mu\nu}=i\,\mathrm{diag}(f^1_{\mu\nu},f^2_{\mu\nu},f^3_{\mu\nu})$ with
$\sum_if^i_{\mu\nu}=0$. The adjoint representation splits into the Cartan
directions, which lie in $\mathfrak h_B$, and three real root planes on which
$\ad_F$ rotates by $f^i-f^j$. Hence, on the physical subspace
$\mathfrak h_B^\perp$,
\begin{align}
 \mathrm{spec}\,K_B^{\rm phys}(x)&=\{\lambda_{12},\lambda_{12},\lambda_{23},
 \lambda_{23},\lambda_{13},\lambda_{13}\},\nonumber\\
 \lambda_{ij}&=\sum_{\mu<\nu}\!\bigl(f^i_{\mu\nu}-f^j_{\mu\nu}\bigr)^2,
\label{eq:spec}
\end{align}
and $\sqrt{\det K^{\rm phys}_B}=\lambda_{12}\lambda_{23}\lambda_{13}$. If some
$\lambda_{ij}$ vanishes, that root plane joins the stabilizer and must not be
counted as a fluctuation mode. For any compact simple $G$ with roots $\Phi$ the
result is $\lambda_\alpha=\sum_{\mu<\nu}\alpha(F_{B,\mu\nu})^2$,
$\alpha\in\Phi^+$, each of multiplicity two; the orbit is the flag manifold
$G/T$ of dimension $2|\Phi^+|$ whenever the span of $\{F_{B,\mu\nu}\}$ contains a
regular element, and $\sum_{\alpha>0}\lambda_\alpha=h^\vee\sum\|F_{B,\mu\nu}\|^2$,
consistent with Eq.~\eqref{eq:sumrule}. Because $K_B$ is algebraic, the
operator spectrum is the essential range of the $\lambda_\alpha(x)$, with lower
edge $\Delta_{\rm alg}=\ess_x\min_\alpha\lambda_\alpha(x)$. This is a gap of the
\emph{Hessian}, not a particle mass: the Gaussian two-point function is
supported at coincident points and carries no correlation length. A strictly
positive $\Delta_{\rm alg}$ also requires, e.g., compact $M$ and
nowhere-degenerate curvature; for finite-action backgrounds on noncompact $M$,
$F_B\to0$ forces $\Delta_{\rm alg}=0$.

\emph{Measure and exact finite-regulator integral.}---The natural measure on
the group variable is normalized Haar measure; in exponential coordinates
$\dd g=J(\epsilon)\dd^8\epsilon$ with
$J=\prod_{\alpha>0}[\sin(\alpha(\epsilon)/2)/(\alpha(\epsilon)/2)]^2$. The
Riemannian volume of the Hessian metric, $\propto\sqrt{\det K^{\rm phys}_B}$, is
a different measure, and neither is a Faddeev--Popov determinant. On a lattice,
with independent site variables $g_n$, the partition function at fixed
background is the exact compact integral
\begin{equation}
 Z[B]=\prod_n I_\kappa(\Theta_n),\qquad
 I_\kappa(\Theta)=\!\int_G\!\dd g\,e^{-\kappa V_\Theta(g)},
\label{eq:I}
\end{equation}
where $\Theta_n=(\Theta_{I}(n))$ are the independent lattice curvature
components, $V_\Theta(g)=\sum_I\|\Ad_{g^{-1}}\Theta_I-\Theta_I\|^2$, and
$\kappa$ absorbs the normalization of the action and the lattice spacing
($\kappa=2a^4$ for the weight $e^{-S}$). The integrand is invariant under the
stabilizer, so the full-group and quotient integrals coincide. Write
$Q=\sum_I\|\Theta_I\|^2$. Four exact facts follow.

(a) \emph{Moments.} For simple $G$, $\Ad$ is irreducible, so
$\int\!\dd g\,\Ad_{g^{-1}}X=0$ and
$\int\!\dd g\,\ip{\Ad_{g^{-1}}X}{Y}\ip{\Ad_{g^{-1}}X'}{Y'}=\tfrac1d\ip X{X'}\ip Y{Y'}$,
$d=\dim G$. Hence $\avg V=2Q$ and
$\mathrm{Var}\,V=\tfrac4d\sum_{I,J}\ip{\Theta_I}{\Theta_J}^2$.

(b) \emph{Bounds.} Since $0\le V\le4Q$, Jensen's inequality gives
$e^{-2\kappa Q}\le I_\kappa\le1$.

(c) \emph{Cumulants.}
\begin{equation}
 \log I_\kappa=-2\kappa Q+\frac{2\kappa^2}{d}\sum_{I,J}\ip{\Theta_I}{\Theta_J}^2+O(\kappa^3),
\label{eq:cum}
\end{equation}
so the leading term is the Yang--Mills density, and the first correction is
quartic in the curvature.

(d) \emph{Closed form for one component.} For a single curvature component
$\Theta=i\,\mathrm{diag}(\theta_1,\dots,\theta_N)$ in $SU(N)$ the integral is of
Harish-Chandra--Itzykson--Zuber type~\cite{HC,IZ},
\begin{equation}
 I_\kappa(\Theta)=e^{-2\kappa\sum_i\theta_i^2}\,
 \frac{\prod_{p=1}^{N-1}p!\ \det\bigl[e^{2\kappa\theta_i\theta_j}\bigr]}
 {(2\kappa)^{N(N-1)/2}\,\Delta(\theta)^2},
\label{eq:HCIZ}
\end{equation}
with $\Delta(\theta)=\prod_{i<j}(\theta_i-\theta_j)$. For several
noncommuting components the exponent contains several pairings with the same
group element, and this is not a single HCIZ integral; character expansions or
numerical group integration are then used.

\emph{Radon--Nikodym relation to Yang--Mills theory.}---Let $\Lambda$ be a
finite hypercubic lattice with link variables $U_l\in G$, symmetric clover
curvature $\Theta_{\mu\nu}(n;U)$, and orbit variables $g_n$. Because each $g_n$
appears in exactly one term, it can be integrated out exactly at fixed $U$
(ultralocality). With no independent gauge action, the induced link measure and
the lattice Yang--Mills measure with the same curvature are
\begin{align}
 \dd\mu_{\Delta F}&=Z_{\Delta F}^{-1}\prod_l\dd U_l\prod_nI_\kappa(\Theta(n;U)),\nonumber\\
 \dd\mu_{\rm YM}&=Z_{\rm YM}^{-1}\prod_l\dd U_l\;e^{-2\kappa\sum_nQ_n(U)}.
\label{eq:measures}
\end{align}
Define $\widetilde I_\kappa=e^{2\kappa Q}I_\kappa$ and
$\varphi_n=\log\widetilde I_\kappa(\Theta(n;U))$. Then the RN derivative is
\begin{equation}
 R_\Lambda=\frac{\dd\mu_{\Delta F}}{\dd\mu_{\rm YM}}
 =\frac{\prod_n\widetilde I_\kappa(\Theta(n;U))}
 {\bigl\langle\prod_n\widetilde I_\kappa\bigr\rangle_{\rm YM}}
 =\frac{e^{\sum_n\varphi_n}}{\bigl\langle e^{\sum_n\varphi_n}\bigr\rangle_{\rm YM}}.
\label{eq:RN}
\end{equation}

\emph{Theorem (finite-regulator RN relation).} For every compact $G$, finite
$\Lambda$, and $\kappa\ge0$: (1) each local factor satisfies
\begin{equation}
 0\le\varphi_n\le2\kappa^2Q_n^2
\label{eq:phibound}
\end{equation}
(the lower bound is (b); the upper bound is Hoeffding's lemma applied to
$V\in[0,4Q]$ with mean $2Q$~\cite{Hoeffding});
(2) $R_\Lambda$ and $R_\Lambda^{-1}$ are strictly positive, gauge invariant, and
bounded, so $\mu_{\Delta F}$ and $\mu_{\rm YM}$ are mutually absolutely
continuous; (3) for every observable $\mathcal O$,
\begin{align}
 \avg{\mathcal O}_{\rm YM}&=\frac{\avg{\mathcal O\,R_\Lambda^{-1}}_{\Delta F}}
 {\avg{R_\Lambda^{-1}}_{\Delta F}},\nonumber\\
 \avg{\mathcal O}_{\Delta F}-\avg{\mathcal O}_{\rm YM}
 &=\frac{\mathrm{Cov}_{\rm YM}(\mathcal O,R_\Lambda)}{\avg{R_\Lambda}_{\rm YM}},
\label{eq:inverse}
\end{align}
and $\int R_\Lambda^{-1}\dd\mu_{\Delta F}=1$, with
$Z_{\Delta F}=Z_{\rm YM}\avg{\prod_n\widetilde I_\kappa}_{\rm YM}$. The proof is
direct: the orbital factors $I_\kappa\widetilde I_\kappa^{-1}=e^{-2\kappa Q}$
cancel exactly.

The content of this relation is structural. The curvature-displacement
measure is the Gibbs measure of lattice Yang--Mills theory with one extra
local, gauge-invariant, curvature-dependent term,
$S_{\rm eff}=\sum_n\bigl[2\kappa Q_n-\varphi_n\bigr]$, and $\varphi_n\ge0$ is
bounded by a \emph{quartic} function of the local curvature. For smooth
backgrounds, $\Theta=a^2F+O(a^3)$ gives $\sum_n\varphi_n\le2\kappa^2a^4\!\int\!\dd^4x\,(\sum_{\mu<\nu}\|F_{\mu\nu}\|^2)^2\to0$, i.e. the
classical scaling of a dimension-eight operator. Thus the Yang--Mills
measure is obtained \emph{exactly} from the ultralocal orbital theory by
reweighting with $R_\Lambda^{-1}$, with no asymptotic assumption. This is the
path we propose: mass-gap and clustering questions for Yang--Mills theory become
questions about the effective background theory $\mu_{\Delta F}$ and about the
uniform behavior of a local, positive cocycle.

We stress what Eq.~\eqref{eq:inverse} does \emph{not} give. At fixed volume the
cocycle is bounded, but the bound $e^{2\kappa^2\sum_nQ_n^2}$ grows with
$|\Lambda|$, so no volume-independent control of connected correlators
follows. The smooth-field estimate is classical: in thermal configurations the
curvature fluctuates at the scale $\kappa Q_n=O(1)$ per site, where $\varphi_n$
is not small, so the cocycle is an extensive reweighting and the irrelevance of
$\varphi_n$ is not established. What is required is (i) uniform-in-volume
integrability of the normalized inverse cocycle, (ii) control of its effect on
the subleading spectrum of the effective transfer operator along a scaling
trajectory $\kappa=\kappa(a)$, and (iii) the identification of the physical
gauge-invariant sector. If a continuum limit exists and the inverse cocycle
preserves a positive lower bound on the physical gap uniformly, the gap
transfers; that hypothesis is exactly the missing estimate and is not proved
here. Normalizing the transfer operator to unit norm does not yield a gap, and
neither does the Hessian gap $\Delta_{\rm alg}$.

\emph{Status of reflection positivity, continuum limit, and mass gap.}---
We summarize what is known, and what this work adds. (1) The Wilson plaquette
weight is reflection positive~\cite{OSeiler}, and lattice Schwinger functions
satisfy the Osterwalder--Schrader (OS) axioms at finite spacing~\cite{OS1,OS2,Seiler}. For the
induced orbital measure with symmetric clover curvature, site-reflection
positivity holds provided the base gauge weight admits a site-reflection-positive
factorization. Link-reflection positivity, needed for a nonnegative
transfer operator $\mathcal T$ and $H=-\log\mathcal T$, is established only
for plaquette-wise central weights with nonnegative character coefficients and
is open for the full clover weight. (2) At fixed spacing, subsequential
infinite-volume limits exist by compactness. The continuum limit of
four-dimensional Yang--Mills theory, and OS reconstruction of its
Schwinger functions, are not established; Bałaban's renormalization-group
analysis controls the ultraviolet problem only in finite volume
\cite{Balaban}, and asymptotic freedom is a perturbative statement~\cite{GW,Politzer}.
(3) A positive mass gap is known at strong coupling by cluster expansions
\cite{OSeiler,Seiler}, not uniformly along a continuum scaling
trajectory. The present paper proves none of these continuum statements, for
either $\mu_{\Delta F}$ or $\mu_{\rm YM}$; what it supplies is the exact
bridge between the two measures, Eq.~\eqref{eq:RN}, so that any such result
for one can be tested against the other.

\emph{Topology.}---A winding-number sector does \emph{not} follow from
Eq.~\eqref{eq:S}. The metric-contracted $\Tr(\Delta F_{\mu\nu}\Delta F^{\mu\nu})$
is not a characteristic density, and ultralocality yields a precise obstruction:
for any smooth $h:M\to SU(3)$ with $h(x)\in H_B(x)$, $S[hg;B]=S[g;B]$. If
$H_B$ contains a fixed $SU(2)$ (e.g.\ $F_B\propto i\,\mathrm{diag}(1,1,-2)$),
$h$ can carry arbitrary degree $w(h)$ on a three-sphere, and $w(hg)=w(h)+w(g)$,
so $S$ does not determine $w$. A winding number appears only under an
additional premise: $M$ compact with $\partial M\cong S^3$; $A$ and $B$ smooth,
not gauge-equivalent in the interior; and $A|_{\partial M}=(B|_{\partial M})^g$
with $g:\partial M\to SU(3)$ of degree $w(g)$ not extendable over $M$
(a large gauge transformation). Then the relative Chern--Simons form gives
$\frac1{8\pi^2}\int_M\Tr(\Delta F\wedge(F_A+F_B))=-w(g) \cong\mathbb Z$~\cite{Bott}), and Cauchy--Schwarz yields the
\emph{inequality}
\begin{multline}
 8\pi^2|w(g)|\le\Bigl(\int_M\sum_{\mu<\nu}\|\Delta F_{\mu\nu}\|^2\Bigr)^{1/2}\\
 \times\Bigl(\int_M\sum_{\mu<\nu}\|F_{A,\mu\nu}+F_{B,\mu\nu}\|^2\Bigr)^{1/2}.
\label{eq:bound}
\end{multline}
Nonzero winding therefore forces nonzero action, but Eq.~\eqref{eq:bound} is not
an identity, and for a gauge relation $A=B^g$ extending over all of $M$ one has
$k_A=k_B$ and no winding charge arises. A Hamiltonian $H=f(W)$ on
$\ell^2(\mathbb Z)$ built from the winding operator has a gap only because $f$
is chosen to give one; it is not the Yang--Mills Hamiltonian, and its gap is
unrelated to $\Delta_{\rm alg}$.

\emph{Conclusion.}---We have established the exact curvature-orbit structure
of the curvature-displacement action: an ultralocal functional on
$H_B\backslash G$ with a positive-semidefinite Hessian whose kernel is the
curvature stabilizer, an explicit root-plane spectrum for every compact simple
group, and an exactly solvable finite-regulator group integral. Its main
consequence is a positive, bounded, local Radon--Nikodym cocycle with
$0\le\log\widetilde I_\kappa\le2\kappa^2Q^2$ that reconstructs the lattice
Yang--Mills measure exactly from the ultralocal orbital theory. Whether this
reformulation yields a positive mass gap in the continuum is reduced to a
specific unproved estimate: uniform-in-volume control of the cocycle along a
scaling trajectory. If a genuine quantum spectrum for the orbit variable
itself is wanted, an additional dynamical term is required, since the
pointwise Hessian produces none.

\begin{acknowledgments}
The author declares no competing interests and acknowledges no financial
support. Generative AI systems (ChatGPT, Claude, Gemini, DeepSeek) were used
for exploratory derivations and numerical cross-checks in locally executed
Python code. All results were verified and are the responsibility of the
author.
\end{acknowledgments}

\end{document}